\documentclass[12pt]{article}
\usepackage{latexsym,amsfonts,amssymb}
\makeatletter
\@addtoreset{equation}{subsection}
\makeatother

\begin{document}
\begin{center}
{\Large \bf Pulsating Massive Star:\\Phase Transitions in Superdense Matter}
\\[1.5cm]
 {\bf Vladimir S.~MASHKEVICH}\footnote {E-mail:
  Vladimir.Mashkevich100@qc.cuny.edu}
\\[1.4cm] {\it Physics Department
 \\ Queens College\\ The City University of New York\\
 65-30 Kissena Boulevard\\ Flushing, New York
 11367-1519} \\[1.4cm] \vskip 1cm

{\large \bf Abstract}
\end{center}

In the case of the energy-momentum tensor related to ``ordinary'' matter (perfect fluid representing spin $1/2$ and $1$ fields), the equations of general relativity result in cosmological and gravitational collapse singularities---due to the fulfilment of both the strong and weak energy conditions. According to great unified theories, in superdense matter (both hot and cold), phase transitions take place, symmetry between the strong and electroweak interactions is restored/broken, and a scalar field is created/annihilated. In the scalar field regime, the strong energy condition is broken, but the weak one holds. However, the continuity condition for the pressure on the surface of a contracting star results in the occurrence of a compensating pseudomatter field, for which both energy conditions are broken. On this basis, for a massive star predisposed to a gravitational collapse, a pulsation dynamics with no singularity is constructed.

\newpage

\section*{Introduction}

In the case of the energy-momentum tensor $T_{\mu}^{\nu}$ related to ``ordinary'' matter (i.e., perfect fluid representing spin $1/2$ and $1$ fields), the equations of general relativity inevitably result in singularities---both cosmological (in the past) and gravitational collapse ones (in the future) [1-3]. The presence of singularities has provoked concern of physicists. Einstein himself had fought against them starting in $1918$ [4].

The inevitability of singularities follows from the singularity theorems [3], which, in turn, are based on the strong and weak energy conditions. Denote by $w_{(\lambda)},\;\lambda=0,1,2,3$, the eigenvalues of the tensor $T_{\mu}^{\nu}$, so that $w_{(0)}=\varrho$ (the energy density) and $w_{(i)}=-p_{(i)},\;i=1,2,3$ (the principal pressures). The energy conditions read: $\varrho+\sum_{i}p_{(i)}\geq 0\;\mathrm{and}\;\varrho+p_{(i)}\geq 0$ (strong); $\varrho\geq 0\;\mathrm{and}\;\varrho+p_{(i)}\geq 0$ (weak). The strong energy condition relates to cosmological singularity, and both energy conditions are related to gravitational collapse.

In the case of ordinary matter, $\varrho\geq 0\;\mathrm{and}\;p_{(i)}\geq 0$, so that both the strong and weak energy conditions are satisfied.

The occurrence of singularities indicates a breakdown of general relativity: its dynamical equations are self-limiting in their scope [5] since they result in singularities, beyond which dynamics cannot be extended.

According to grand unified theories, in superdense matter (both hot and cold), phase transitions take place, after which symmetry between the strong and electroweak interactions is restored/broken and a scalar field $\varphi$ (``extraordinary'' matter) is created/annihilated [6]. In the scalar field regime, $\varrho=(1/2)\dot{\varphi}^{2}+V(\varphi),\;p_{(i)}=p=(1/2)\dot{\varphi}^{2}-V(\varphi)$ [$^{.}=\mathrm{d}/\mathrm{d}t,\;V(\varphi)$ is the potential], so that $\varrho>0,\;\varrho+p_{(i)}\geq 0,\;\varrho+\sum_{i}p_{(i)}=2[\dot{\varphi}^{2}-V(\varphi)]$. Thus, the strong energy condition may be broken. Starting from this circumstance, in [7] a theory of deflation-inflation for the universe has been advanced, which results in the contracting-expanding universe with no singularity.

On the other hand, for the scalar field, the weak energy condition holds, so that it seems that gravitational collapse (i.e., singularity) is inevitable. However, the existence of the boundary surface of a contracting star results in the occurrence of an external pseudomatter field [8], which compensates for the discontinuity of the scalar field pressure on the boundary surface. Thus, extraordinary matter is comprised of the interior scalar and exterior pseudomatter fields. (The latter represents dark matter.)

For the pseudomatter field, the result is this: $p_{(\mathrm{normal})}<0,\;\varrho<0$, so that both energy conditions are broken, which makes it possible to avoid singularity.

In the present paper, for a massive star predisposed to a gravitational collapse, a  pulsation dynamics with no singularity is constructed.
\newpage

\section{Basic equations}

\subsection{The extended Einstein equations. Matter and pseudomatter}

In the extended Einstein equations [8]
\begin{equation} 
G_{\mu}^{\nu}-8\pi\varkappa T_{\mu}^{\nu}=0,\quad\mu,\nu=0,1,2,3
\end{equation}
($\hbar=c=1,\;\varkappa=t_{\mathrm{Planck}}^{2}$ is the gravitational constant) the energy-momentum tensor involves both matter (m) and pseudomatter (ps), which represents dark matter:
\begin{equation} 
T_{\mu}^{\nu}=T_{\mathrm{m}}{}_{\mu}^{\nu}+T_{\mathrm{ps}}{}^{\nu}_{\mu}
\end{equation}
$T_{\mathrm{ps}}{}^{\nu}_{\mu}$ compensates for the discontinuity of $T_{\mathrm{m}}{}_{\mu}^{\nu}$ in time and space.

\subsection{Spherically symmetric star. Dimensionless quantities}

We consider a spherically symmetric star; denote by $a$ its radius and introduce dimensionless quantities (the subscript dim stands for dimensional):

time: $t=t_{\mathrm{dim}}/a$;$\quad$ radial coordinate: $r=r_{\mathrm{dim}}/a$;

radius of sphere with radial coordinate $r$: $K(r,t)=K_{\mathrm{dim}}(r,t)/a$;

metric: $\mathrm{d}s^{2}=\mathrm{d}s^{2}_{\mathrm{dim}}/a^{2}$;

the Einstein tensor: $G_{\mu}^{\nu}=a^{2}G_{\mu\,\mathrm{dim}}^{\nu}$;$\quad$the energy-momentum tensor: $T_{\mu}^{\nu}=a^{2}\varkappa T_{\mu\,\mathrm{dim}}^{\nu} $;

energy density: $\varrho=a^{2}\varkappa\rho_{\mathrm{dim}}$;$\quad$pressure: $p=a^{2}\varkappa p_{\mathrm{dim}}$;

scalar field: $\varphi=\sqrt{a^{2}\varkappa}\varphi_{\mathrm{dim}}$;$\quad$ potential: $V=a^{2}\varkappa V_{\mathrm{dim}}$.

\subsection{Metric, the Einstein equations and tensor components}

In the case of spherical symmetry, in the synchronous reference frame, metric is of the form [2]:
\begin{equation} 
\mathrm{d}s^{2}=\mathrm{d}t^{2}-\mathrm{e}^{\lambda(r,t)}\mathrm{d}r^{2}-
K^{2}(r,t)(\mathrm{d}\theta^{2}+\sin^{2}\theta\,\mathrm{d}\phi^{2})
\end{equation}

Again, in the case of spherical symmetry, the nonvanishing tensor components are
\begin{equation} 
X_{0}^{0},\;X_{1}^{1},\;X_{2}^{2}=X_{3}^{3},\;X_{0}^{1},\;X_{1}^{0};\quad X=G,T_{\mathrm{m}},T_{\mathrm{ps}}
\end{equation}
where
\begin{equation} 
x^{0}=t,\;x^{1}=r,\;x^{2}=\theta,\;x^{3}=\phi
\end{equation}

Thus, (1.1.1) in the dimensionless form reduces to
\begin{equation} 
G_{1}^{1}-8\pi T_{1}^{1}=0
\end{equation}
\begin{equation} 
G_{2}^{2}-8\pi T_{2}^{2}=0
\end{equation}
\begin{equation} 
G_{0}^{0}-8\pi T_{0}^{0}=0
\end{equation}
\begin{equation} 
G_{1}^{0}-8\pi T_{1}^{0}=0
\end{equation}
Now [2], ($^{.}=\partial/\partial t,\;'=\partial/\partial r$)
\begin{equation} 
G_{1}^{1}=\frac{2\ddot{K}}{K}+\frac{\dot{K}^{2}}{K^{2}}+
\frac{1}{K^{2}}-\frac{(K')^{2}}{K^{2}}\mathrm{e}^{-\lambda}
\end{equation}
\begin{equation} 
G_{2}^{2}=\frac{\ddot{K}}{K}+\frac{\ddot{\lambda}}{2}+\frac{\dot{\lambda}^{2}}{4}+
\frac{\dot{K}\dot{\lambda}}{2K}-\frac{1}{2}\left(\frac{2K''}{K}-
\frac{K'\lambda'}{K}\right)\mathrm{e}^{-\lambda}
\end{equation}
\begin{equation} 
G_{0}^{0}=\frac{\dot{K}^{2}}{K^{2}}-\frac{\dot{K}\dot{\lambda}}{K}+\frac{1}{K^{2}}-
\left(\frac{2K''}{K}-
\frac{K'\lambda'}{K}\right)\mathrm{e}^{-\lambda}-\frac{(K')^{2}}{K^{2}}\mathrm{e}^{-\lambda}
\end{equation}
\begin{equation} 
G_{1}^{0}=\frac{2\dot{K}'}{K}-\frac{\dot{\lambda}K'}{K}
\end{equation}

\subsection{Matter and pseudomatter equations}

Matter and pseudomatter equations are of the form
\begin{equation} 
T_{\mathrm{z}}{}_{\mu;\lambda}^{\lambda}=0,\quad \mathrm{z=m,ps}
\end{equation}
i.e.,
\begin{equation} 
T_{\mathrm{z}}{}_{\mu,\lambda}^{\lambda}+\Gamma^{\lambda}_{\lambda\sigma}
T_{\mathrm{z}}{}_{\mu}^{\sigma}-\Gamma^{\sigma}_{\mu\lambda}
T_{\mathrm{z}}{}_{\sigma}^{\lambda}=0
\end{equation}
The nonvanishing Christoffel symbols are [2]
\begin{eqnarray}
\Gamma_{11}^{1}&=&\lambda'/\lambda\qquad\Gamma_{22}^{1}=-\mathrm{e}^{-\lambda}KK'\qquad
\Gamma_{33}^{1}=-\mathrm{e}^{-\lambda}KK'\sin^{2}\theta\qquad \Gamma_{10}^{1}=\dot{\lambda}/2
\nonumber\\
\Gamma_{12}^{2}&=&K'/K\qquad\Gamma_{20}^{2}=\dot{K}/K\qquad
\Gamma_{33}^{2}=-\sin\theta\cos\theta\nonumber\\
\Gamma_{13}^{3}&=&K'/K\qquad\Gamma_{30}^{3}=\dot{K}/K\qquad
\Gamma_{23}^{3}=\cot\theta\nonumber\\
\Gamma_{11}^{0}&=&\dot{\lambda}\mathrm{e}^{\lambda}/2\qquad\Gamma_{22}^{0}=K\dot{K}\qquad
\Gamma_{33}^{0}=K\dot{K}\sin^{2}\theta
\end{eqnarray}

As a consequence of (1.4.1), the dynamical equations are (1.3.8), (1.3.9), whereas (1.3.10), (1.3.11) are constraints on initial conditions.

\section{Setting of a problem}

\subsection{Metric singularity}

The definition of the singularity of metric is this:
\begin{equation} 
K(r,t)\rightarrow 0\quad\mathrm{and}\quad\mathrm{e}^{\lambda(r,t)}\rightarrow 0\quad
 \mathrm{for}\;t\rightarrow t_{\mathrm{sing}}<\infty
\end{equation}
The existence of singularities is the consequence of the fulfilment of the energy conditions [3].

A gravitational collapse results in the singularity (2.1.1). There arises the problem of a mass point with a metric singularity at the point (rather than a mass point in a regular metric field). For this problem, the equations of general relativity, specifically the geodesic equation, fail. Thus, it is impossible to construct a consistent dynamics for the universe with singularities.

In this connection, we quote Penrose [5]:

``This is not to say that some mathematically precise concept of `singularity' should now form part of our description of physical geometry---though much elegant work has been done in this direction in recent years. Rather, it seems to be that the very notion of spacetime geometry, and consequently the physical laws as we presently understand them, are limited in their scope. Indeed, these laws are even $\it{self}$-limiting, as the singularity theorems ... seem to show. ...There is a need for new laws in any case...''

\subsection{The energy conditions}

Denote by $w$ the eigenvalues of the energy-momentum tensor, i.e., solutions to the equations

\begin{equation} 
T_{\mu}^{\nu}v_{\nu}=wv_{\mu},\quad w=w_{(\lambda)},\;\;\lambda=0,1,2,3,\;\;v\neq 0
\end{equation}
The standard notation is this:
\begin{equation} 
w_{(0)}=\varrho,\quad w_{(i)}=-p_{(i)},\;\;i=1,2,3
\end{equation}
The eigenvalue $\varrho$ may be interpreted as the density, and the eigenvalues $p_{(i)}$ are called the principal pressures [3].

The strong energy condition is
\begin{equation} 
\varrho+\sum\limits_{i=1}^{3}p_{(i)}\geq 0\quad\mathrm{and}\quad \varrho+p_{(i)}\geq 0,\;\;
i=1,2,3
\end{equation}

The weak energy condition is
\begin{equation} 
\varrho\geq 0\quad\mathrm{and}\quad \varrho+p_{(i)}\geq 0,\;\;i=1,2,3
\end{equation}

The fulfilment of the strong energy condition results in a cosmological singularity. The fulfilment of the weak or strong energy condition results in a gravitational collapse singularity [3].

The case of cosmology has been considered in [7]. In the present paper, we are interested in the case of a star. Therefore our problem is, first of all, to determine if both the weak and strong energy conditions may be broken.

\subsection{Ordinary and extraordinary matter}

With respect to density and pressure, there are ordinary matter (perfect fluid representing spin $1/2$ and $1$ fields) and two kinds of extraordinary matter: scalar field and pseudomatter.

$\it{Ordinary\; matter}$

In the case of ordinary matter,
\begin{equation} 
\varrho_{\mathrm{ord}}\geq 0,\quad p_{\mathrm{ord}}\geq 0,\;\;i=1,2,3
\end{equation}
so that both the strong and weak energy conditions are fulfilled.
\newpage
$\it Scalar\, field$

According to grand unified theories, in superdense matter (both hot and cold), phase transitions occur, after which symmetry between the strong and electroweak interactions is restored/broken, with the result that a scalar field $\varphi$ (extraordinary matter) is created/annihilated [6]. The energy density and pressure of a spatially homogeneous scalar field $\varphi(t)$ with potential $V(\varphi)$ are of the form
\begin{equation} 
\varrho_{sc}=\frac{1}{2}\dot{\varphi}^{2}+V(\varphi),\quad p_{\mathrm{sc}(i)}=
p_{\mathrm{sc}}=\frac{1}{2}\dot{\varphi}^{2}-V(\varphi),\;\;i=1,2,3
\end{equation}
We have
\begin{equation} 
\varrho_{sc}> 0
\end{equation}
\begin{equation} 
\varrho_{sc}+p_{\mathrm{sc}(i)}=\dot{\varphi}^{2}\geq 0,\;\;i=1,2,3
\end{equation}
\begin{equation} 
\varrho_{sc}+\sum\limits_{i}p_{\mathrm{sc}(i)}=\varrho_{sc}+3p_{\mathrm{sc}}=
2[\dot{\varphi}^{2}-V(\varphi)]
\end{equation}
Thus, the strong energy condition may be broken, but the weak energy condition is fulfilled.

$\it{Pseudomatter}$

It is the behavior of the energy conditions for pseudomatter that is the crux of the singularity problem.

\subsection{The problem}

Now the problem is this: To determine if due to the involvement of pseudomatter both the strong and weak energy conditions may be broken and, if so, to construct a pulsation dynamics of a star.

\section{Model}

\subsection{Interior and exterior}

Assume that in the interior of the star there is only matter and in the exterior only pseudomatter:
\begin{equation} 
\mathrm{interior}:\;\;0\leq r< 1,\;\;T_{\mathrm{ps}}{}_{\mu}^{\nu}=0,\;\;T_{\mu}^{\nu}=
T_{\mathrm{m}}{}_{\mu}^{\nu}
\end{equation}
\begin{equation} 
\mathrm{exterior}:\;\;1\leq r<\infty,\;\;T_{\mathrm{m}}{}_{\mu}^{\nu}=0,\;\;T_{\mu}^{\nu}=
T_{\mathrm{ps}}{}_{\mu}^{\nu}
\end{equation}

\subsection{The matter energy-momentum tensor and matter equations}

Let
\begin{equation} 
T_{\mathrm{m}}{}_{0}^{0}=:\varrho_{\mathrm{m}},\;\;T_{\mathrm{m}}{}_{1}^{1}=
T_{\mathrm{m}}{}_{2}^{2}=T_{\mathrm{m}}{}_{3}^{3}=:-p_{\mathrm{m}}
\end{equation}
and\begin{equation} 
T_{\mathrm{m}}{}_{1}^{0}=T_{\mathrm{m}}{}_{0}^{1}=0
\end{equation}
This corresponds to the perfect fluid:
\begin{equation} 
T_{\mathrm{m}}{}_{\mu}^{\nu}=(\varrho_{\mathrm{m}}+p_{\mathrm{m}})
u_{\mathrm{m\mu}}u_{\mathrm{m}}{}^{\nu}-p\delta_{\mu}^{\nu}
\end{equation}
with
\begin{equation} 
u_{\mathrm{m}}{}^{i}=0,\;\;u_{\mathrm{m}0}u_{\mathrm{m}}{}^{0}=1
\end{equation}

Now matter equations (Subsection 1.4) reduce to
\begin{equation} 
T_{\mathrm{m}2}{}_{;\lambda}^{\lambda}=T_{\mathrm{m}3}{}_{;\lambda}^{\lambda}
\equiv 0
\end{equation}
\begin{equation} 
T_{\mathrm{m}}{}_{,0}^{0}+(\Gamma_{10}^{1}+2\Gamma_{20}^{2})
(T_{\mathrm{m}}{}_{0}^{0}-T_{\mathrm{m}}{}_{1}^{1})=0
\end{equation}
\begin{equation} 
T_{\mathrm{m}1}{}_{,1}^{1}=0
\end{equation}
Thus the pressure is spatially homogeneous:
\begin{equation} 
p'_{\mathrm{m}}=0,\;\;p_{\mathrm{m}}=p_{\mathrm{m}}(t)
\end{equation}

\subsection{Homogeneous density, interior metric, \\the interior Einstein equations, and matter equation}

In view of (3.2.8), put
\begin{equation} %
\varrho_{\mathrm{m}}'=0,\;\;\varrho_{\mathrm{m}}=\varrho_{\mathrm{m}}(t)
\end{equation}
Under the homogeneity conditions (3.3.1), (3.2.8), we have in the interior
\begin{equation} 
K_{\mathrm{int}}(r,t)=R(t)r,\quad (\mathrm{e}^{\lambda})_{\mathrm{int}}=
\frac{R^{2}(t)}{1-r^{2}},\quad 0\leq r<1
\end{equation}
and
\begin{equation} 
(\mathrm{d}s^{2})_{\mathrm{int}}=\mathrm{d}t^{2}-R^{2}(t)\left[
\frac{\mathrm{d}r^{2}}{1-r^{2}}+r^{2}(\mathrm{d}\theta^{2}+
\sin^{2}\theta\,\mathrm{d}\phi^{2})\right],\quad 0\leq r<1
\end{equation}

Now the interior Einstein equations reduce to
\begin{equation} 
\frac{2\ddot{R}}{R}+\frac{\dot{R}^{2}}{R^{2}}+\frac{1}{R^{2}}+
8\pi p_{\mathrm{m}}=0
\end{equation}
\begin{equation} 
3\left(\frac{\dot{R}^{2}}{R^{2}}+\frac{1}{R^{2}}\right)-8\pi\varrho_{\mathrm{m}}=0
\end{equation}
and the matter equations reduce to
\begin{equation} 
\dot{\varrho}_{\mathrm{m}}R+3(\varrho_{\mathrm{m}}+p_{\mathrm{m}})\dot{R}=0
\end{equation}
Dynamical equations are (3.3.4), (3.3.6), and (3.3.5) is a constraint on initial conditions.

Without pseudomatter we would have
\begin{equation} 
p_{\mathrm{m}}(1,t)=0
\end{equation}
which would result in
\begin{equation} 
p_{\mathrm{m}}(r,t)=0
\end{equation}
Later on, we will put
\begin{equation} 
p_{\mathrm{ord}}=0
\end{equation}

\subsection{Relation between matter density and pressure}

For ordinary matter, the relation between density and pressure is given by the state equation:
\begin{equation} 
p_{\mathrm{ord}}=p_{\mathrm{ord}}(\varrho_{\mathrm{ord}})
\end{equation}

For scalar field, density and pressure are related via the field:
\begin{equation} 
\varrho_{\mathrm{sc}}=\frac{1}{2}\dot{\varphi}^{2}+V(\varphi),\quad
p_{\mathrm{sc}}=\frac{1}{2}\dot{\varphi}^{2}-V(\varphi)
\end{equation}
The matter equation (3.3.6) reduces to
\begin{equation} 
\dot{\varphi}\left[R\ddot{\varphi}+3\dot{R}\dot{\varphi}+
R\frac{\mathrm{d}V}{\mathrm{d}\varphi}\right]=0
\end{equation}
i.e., either
\begin{equation} 
\dot{\varphi}=0
\end{equation}
or
\begin{equation} 
R\ddot{\varphi}+3\dot{R}\dot{\varphi}+
R\frac{\mathrm{d}V}{\mathrm{d}\varphi}=0
\end{equation}
If both (3.4.5) and (3.4.4) hold, then
\begin{equation} 
\frac{\mathrm{d}V}{\mathrm{d}\varphi}=0
\end{equation}
In the case of $\dot{\varphi}=0$,
\begin{equation} 
-p_{\mathrm{sc}}=\varrho_{\mathrm{sc}}=V(\varphi)=\mathrm{const}
\end{equation}

\subsection{The pseudomatter energy-momentum tensor,\\ pseudomatter equations, and the exterior Einstein equations}

Taking into account matching conditions at $r=1$ [2] and (3.2.1), (3.2.2), (3.2.3), (1.4.1), we put
\begin{equation} 
T_{\mathrm{ps}}{}_{2}^{2}=T_{\mathrm{ps}}{}_{3}^{3}=0,\quad T_{\mathrm{ps}}{}_{1}^{0}=
T_{\mathrm{ps}}{}_{0}^{1}=0
\end{equation}
but must retain
\begin{equation} 
T_{\mathrm{ps}}{}_{1}^{1}=:-p_{\mathrm{ps}1},\quad T_{\mathrm{ps}}{}_{0}^{0}
=:\varrho_{\mathrm{ps}}
\end{equation}

The pseudomatter equations take the form
\begin{equation} 
T_{\mathrm{ps}0}{}_{,0}^{0}+\left(\frac{\dot{\lambda}}{2}+
2\frac{\dot{K}}{K}\right)T_{\mathrm{ps}}{}_{0}^{0}-\frac{\dot{\lambda}}{2}
T_{\mathrm{ps}}{}_{1}^{1}=0
\end{equation}
\begin{equation} 
T_{\mathrm{ps}1}{}_{,1}^{1}+2\frac{K'}{K}T_{\mathrm{ps}}{}_{1}^{1}=0
\end{equation}
and the exterior Einstein equations are these:
\newpage
\begin{equation} 
G_{1}^{1}-8\pi T_{\mathrm{ps}}{}_{1}^{1}=0
\end{equation}
\begin{equation} 
G_{2}^{2}=0
\end{equation}
\begin{equation} 
G_{0}^{0}-8\pi T_{\mathrm{ps}}{}_{0}^{0}=0
\end{equation}
\begin{equation} 
G_{1}^{0}=0
\end{equation}
with $G_{\mu}^{\nu}$ given by (1.3.8)--(1.3.11).

(3.5.3)--(3.5.6) are dynamical equations; (3.5.7), (3.5.8) are constraints on initial conditions.

\subsection{Matching conditions at $r=1$}

Matching conditions at $r=1$ are these [2]:
\begin{equation} 
p_{\mathrm{ps}1}(1,t)=p_{\mathrm{m}}(1,t)
\end{equation}
\begin{equation} 
K_{\mathrm{ext}}(1,t)=K_{\mathrm{int}}(1,t)=R(t)
\end{equation}

From (3.6.1), (3.6.2), (1.3.4), (1.3.8) follows
\begin{equation} 
[(K')^{2}\mathrm{e}^{-\lambda}]_{\mathrm{ext}}(1,t)=
[(K')^{2}\mathrm{e}^{-\lambda}]_{\mathrm{int}}(1,t)
\end{equation}
Now,
\begin{equation} 
[(K')^{2}\mathrm{e}^{-\lambda}]'=K'[(2K''-K'\lambda')\mathrm{e}^{-\lambda}]
\end{equation}
so that the expression $(2K''-K'\lambda')$ involved in (1.3.9), (1.3.10) does not contain $\delta(r-1)$.

Next,
\begin{equation} 
[(K')^{2}\mathrm{e}^{-\lambda}]^{.}=K'\mathrm{e}^{-\lambda}
[2\dot{K}'-\dot{\lambda}K']
\end{equation}
which, with (1.3.11), (3.5.8), results in
\begin{equation} 
[(K')^{2}\mathrm{e}^{-\lambda}]^{.}=0
\end{equation}

Again,
\begin{equation} 
[(K')^{2}\mathrm{e}^{-\lambda}]_{\mathrm{int}}=1-r^{2},\quad
[(K')^{2}\mathrm{e}^{-\lambda}]_{\mathrm{int}}(1,t)=0
\end{equation}
Thus,
\begin{equation} 
[(K')^{2}\mathrm{e}^{-\lambda}]_{\mathrm{ext}}(1,t)=0
\end{equation}
so that
\begin{equation} 
K'_{\mathrm{ext}}(1,t)=0
\end{equation}

\subsection{Pseudomatter pressure and density}

The equation for $p_{\mathrm{ps}1}$ is (3.5.4):
\begin{equation} 
p'_{\mathrm{ps}1}+2\frac{K'}{K}p_{\mathrm{ps}1}=0
\end{equation}
whence
\begin{equation} 
p_{\mathrm{ps}1}K^{2}=f(t),\quad p_{\mathrm{ps}1}(r,t)=\frac{f(t)}{K^{2}(r,t)}
\end{equation}
From (3.6.1) follows
\begin{equation} 
\frac{f(t)}{K^{2}(1,t)}=p_{\mathrm{m}}(1,t)
\end{equation}
Thus, in view of (3.6.2),
\begin{equation} 
p_{\mathrm{ps}1}(r,t)=p_{\mathrm{m}}(1,t)\frac{R^{2}(t)}{K^{2}(r,t)}=
p_{\mathrm{m}}(t)\frac{R^{2}(t)}{K^{2}(r,t)}
\end{equation}
Note: as long as
\begin{equation} 
\lim_{r\rightarrow 0}K(r,t)=0
\end{equation}
$(p_{\mathrm{ps}1})_{\mathrm{int}}$ is inadmissible. So
\begin{equation} 
p_{\mathrm{ps}1}(r,t)=p_{\mathrm{m}}(t)\frac{R^{2}(t)}{K^{2}_{\mathrm{ext}}(r,t)}
\end{equation}

Next, the equation for $\varrho_{\mathrm{ps}}$ is (3.5.3):
\begin{equation} 
\dot{\varrho}_{\mathrm{ps}}+\left(\frac{\dot{\lambda}}{2}+
2\frac{\dot{K}}{K}\right)\varrho_{\mathrm{ps}}+\frac{\dot{\lambda}}{2}p_{\mathrm{ps}1}=0
\end{equation}
Denote by $t_{\mathrm{b-r}}$ the instant of the symmetry broken-restored transition and by $t_{\mathrm{r-b}}$ the instant of the reverse transition (if the latter occurs, i.e., if there is no singularity). Put
\begin{equation} 
\varrho_{\mathrm{ps}}(t_{\mathrm{b-r}})=0
\end{equation}
then
\begin{equation} 
\varrho_{\mathrm{ps}}(r,t)=-\frac{1}{K^{2}(r,t)\mathrm{e}^{\lambda(r,t)/2}}
\int\limits_{t_{\mathrm{b-r}}}^{t}\frac{\partial\mathrm{e}^{\lambda(r,t)/2}}{\partial t}
p_{\mathrm{m}}(t)R^{2}(t)\mathrm{d}t
\end{equation}

\subsection{Breakdown of the energy conditions}

Assume that there is no singularity. Then for
\begin{equation} 
t_{\mathrm{b-r}}\leq t\leq t_{\mathrm{r-b}}
\end{equation}
we have
\begin{equation} 
p_{\mathrm{m}}=p_{\mathrm{sc}}=\frac{1}{2}\dot{\varphi}^{2}-V(\varphi)
\end{equation}
Let
\begin{equation} 
p_{\mathrm{sc}}<0
\end{equation}
then, by (3.7.6),
\begin{equation} 
p_{\mathrm{ps}1}<0
\end{equation}

Introduce
\begin{equation} 
t_{0}=\frac{t_{\mathrm{b-r}}+t_{\mathrm{rb}}}{2}
\end{equation}
We have
\begin{equation} 
\frac{\partial\mathrm{e}^{\lambda/2}}{\partial t}(r,t_{0})=0,\quad
\frac{\partial\mathrm{e}^{\lambda/2}}{\partial t}(r,t)\lessgtr 0
\;\;\mathrm{for}\;t\lessgtr t_{0}
\end{equation}
Thus
\begin{equation} 
\varrho_{\mathrm{ps}}\leq 0\;\;\mathrm{for}\;t\leq t_{\mathrm{r-b}}
\end{equation}

We have obtained the following results: for $t$ given by (3.8.1)
\begin{equation} 
\varrho_{\mathrm{ps}}+\sum\limits_{i}p_{\mathrm{ps}i}<0\quad
\mathrm{and}\quad \varrho_{\mathrm{ps}}+p_{\mathrm{ps}1}<0
\end{equation}
\begin{equation} 
\varrho_{\mathrm{ps}}\leq 0\quad \mathrm{and}\quad \varrho_{\mathrm{ps}}+p_{\mathrm{ps}1}<0
\end{equation}
so that both the strong and weak energy conditions are broken.

In the interior, the strong energy condition is broken as long as
\begin{equation} 
\varrho_{\mathrm{sc}}+\sum\limits_{i}p_{\mathrm{sc}i}=\varrho_{\mathrm{sc}}+
3p_{\mathrm{sc}}=2[\dot{\varphi}^{2}-V(\varphi)]<0
\end{equation}
For the metric (3.3.3), it is the breakdown only of the strong energy condition that is essential.

Thus, the energy conditions are broken in the whole spacetime, and our assumption is justified.

Now we may construct a pulsation dynamics, i.e., one free of singularity.

\section{Pulsation dynamics}

\subsection{Broken symmetry (ordinary matter) stage}

Consider a cycle of a pulsation dynamics. Put
\begin{equation} 
-t_{\mathrm{initial}}=t_{\mathrm{final}}>0
\end{equation}
in which case
\begin{equation} 
-t_{\mathrm{b-r}}=t_{\mathrm{r-b}}>0,\quad t_{0}=\frac{t_{\mathrm{b-r}}+
t_{\mathrm{r-b}}}{2}=0
\end{equation}
The broken symmetry, i.e., ordinary matter stage takes place for
\begin{equation} 
t\in[t_{\mathrm{initial}},t_{\mathrm{b-r}})\bigcup(t_{\mathrm{r-b}},t_{\mathrm{final}}]
\end{equation}
with
\begin{equation} 
\dot{X}(t_{\mathrm{initial}})=\dot{X}(t_{\mathrm{final}})=0,\quad
X=K,\lambda
\end{equation}

$\it{Interior}$

The dynamical equations are:
\begin{equation} 
\frac{2\ddot{R}}{R}+\frac{\dot{R}^{2}}{R^{2}}+\frac{1}{R^{2}}+8\pi p_{\mathrm{ord}}=0
\end{equation}
\begin{equation} 
\dot{\varrho}_{\mathrm{ord}}R+3(\varrho_{\mathrm{ord}}+p_{\mathrm{ord}})\dot{R}=0
\end{equation}
with
\begin{equation} 
p_{\mathrm{ord}}=p_{\mathrm{ord}}(\varrho_{\mathrm{ord}})
\end{equation}
and the constraint
\begin{equation} 
3\left(\frac{\dot{R}^{2}}{R^{2}}+\frac{1}{R^{ 2}}\right)-8\pi \varrho_{\mathrm{ord}}=0
\end{equation}

$\it{Exterior}$

The dynamical equations are:
\begin{equation} 
G_{1}^{1}+8\pi p_{\mathrm{ord}}(t)\frac{R^{2}(t)}{K^{2}(r,t)}=0
\end{equation}
\begin{equation} 
G_{2}^{2}=0
\end{equation}
and the constraints are:
\begin{equation} 
G_{0}^{0}-8\pi\rho_{\mathrm{ps}}=0
\end{equation}
\begin{equation} 
G_{1}^{0}=0
\end{equation}
[$G_{\mu}^{\nu}$ are given by (1.3.8)--(1.3.11)].

In accordance with (3.7.8), we put

\begin{equation} 
p_{\mathrm{ord}}=0,\quad p_{\mathrm{ps}}=0,\;\;\varrho_{\mathrm{ps}}=0
\end{equation}
In this case, there exists an exact, well known solution to the above equations [2].

$\it{Matching}$

Matching conditions at $r=1$ are determined in Subsection 3.6.

\subsection{Restored symmetry (scalar field) stage}

The restored symmetry, i.e., scalar field and pseudomatter stage takes place for
\begin{equation} 
t\in [t_{\mathrm{b-r}},t_{\mathrm{r-b}}]
\end{equation}

$\it{Interior}$

The dynamical equations are:
\begin{equation} 
\frac{2\ddot{R}}{R}+\frac{\dot{R}^{2}}{R^{2}}+\frac{1}{R^{2}}+8\pi
\left[\frac{1}{2}\dot{\varphi}^{2}-V(\varphi)\right]=0
\end{equation}
\begin{equation} 
\dot{\varphi}=0\quad \mathrm{or}\quad R\ddot{\varphi}+3\dot{R}\dot{\varphi}+
R\frac{\mathrm{d}V}{\mathrm{d}\varphi}=0
\end{equation}
with the constraint
\begin{equation} 
3\left(\frac{\dot{R}^{2}}{R^{2}}+\frac{1}{R^{ 2}}\right)-8\pi
\left[\frac{1}{2}\dot{\varphi}^{2}+V(\varphi)\right]=0
\end{equation}

From (4.2.2), (4.2.4) follows
\begin{equation} 
\ddot{R}-\Gamma^{2}R=0
\end{equation}
where
\begin{equation} 
\Gamma^{2}=\frac{8\pi}{3}[V(\varphi)-\dot{\varphi}^{2}]
\end{equation}
The inequality
\begin{equation} 
\Gamma^{2}>0\quad \mathrm{for} \;\;t \;(4.2.1)
\end{equation}
must hold.

We have
\begin{equation} 
\dot{R}(0)=0,\quad\dot{\varphi}(0)=0
\end{equation}
and
\begin{equation} 
R(0)=\frac{1}{\Gamma(0)}=\sqrt{\frac{3}{8\pi V(\varphi(0))}}
\end{equation}

$\it{Exterior}$

The dynamical equations are:
\begin{equation} 
G_{1}^{1}+8\pi\left[\frac{1}{2}\dot{\varphi}^{2}-V(\varphi)\right]
\frac{R^{2}}{K^{2}}=0
\end{equation}
\begin{equation} 
G_{2}^{2}=0
\end{equation}
and the constraints are:
\begin{equation} 
G_{0}^{0}-8\pi\varrho_{\mathrm{ps}}
\end{equation}
\begin{equation} 
G_{1}^{0}=0
\end{equation}
with $G_{\mu}^{\nu}$ given by (1.3.8)--(1.3.11) and $\varrho_{\mathrm{ps}}$ by (3.7.9).

$\it{Matching}$

Matching conditions at $r=1$ are determined in Subsection 3.6.

\subsection{Matching conditions at phase transitions}

Matching conditions at the phase transitions ($t=t_{\mathrm{b-r}},t_{\mathrm{r-b}}$) are these:
\begin{equation} 
\varrho_{\mathrm{ord}}(t_{\mathrm{b-r/r-b}})=\varrho_{\mathrm{sc}}(t_{\mathrm{b-r/r-b}})
\end{equation}
\begin{equation} 
R(t_{\mathrm{b-r/r-b}}+0)=R(t_{\mathrm{b-r/r-b}}-0)
\end{equation}
\begin{equation} 
\dot{R}(t_{\mathrm{b-r/r-b}}+0)=\dot{R}(t_{\mathrm{b-r/r-b}}-0)
\end{equation}

\subsection{The simplest model: Constant scalar field}

Consider the simplest model:
\begin{equation} 
\dot{\varphi}=0,\quad\varphi=\mathrm{const}
\end{equation}
in the restored symmetry stage.

We have
\begin{equation} 
\ddot{R}-\Gamma R=0
\end{equation}
with
\begin{equation} 
\Gamma=\sqrt{\frac{8\pi V(\varphi)}{3}}=\mathrm{const}
\end{equation}
and the initial conditions
\begin{equation} 
R(0)=\frac{1}{\Gamma},\qquad \dot{R}(0)=0
\end{equation}
The solution is this:
\begin{equation} 
R(t)=\frac{1}{\Gamma}\cosh\Gamma t
\end{equation}
or, in the dimensional quantities,
\begin{equation} 
R_{\mathrm{dim}}=\frac{1}{\Gamma_{\mathrm{dim}}}\cosh\Gamma_{\mathrm{dim}} t_{\mathrm{dim}}
\end{equation}
\begin{equation} 
\Gamma_{\mathrm{dim}}=\sqrt{\frac{8\pi\varkappa V(\varphi)}{3}},\quad
V_{\mathrm{dim}}=\frac{1}{a^{2}\varkappa}V
\end{equation}

\subsection{Pulsation dynamics stages (one cycle)}

The pulsation dynamics stages for one cycle are these:

$t_{\mathrm{initial}}:\;\;R=R_{\mathrm{max}},\;\dot{R}=0,\;\ddot{R}<0$

$(t_{\mathrm{initial}},t_{\mathrm{b-r}}):\;\;\dot{R}<0,\ddot{R}<0$

$(t_{\mathrm{b-r}},0):\;\;\dot{R}<0,\ddot{R}>0$

$t=0:\;\;R=R_{\mathrm{min}},\dot{R}=0,\ddot{R}>0$

$(0,t_{\mathrm{r-b}}):\;\;\dot{R}>0,\ddot{R}>0$

$(t_{\mathrm{r-b}},t_{\mathrm{final}}):\;\;\dot{R}>0,\ddot{R}<0$

$t_{\mathrm{final}}:\;\;R=R_{\mathrm{max}},\dot{R}=0,\ddot{R}<0$

\section*{Acknowledgments}

I would like to thank Alex A. Lisyansky for support and Stefan V.
Mashkevich for helpful discussions.


\begin{thebibliography}{99}

\bibitem{1}L.D. Landau, E.M. Lifshitz, The Classical Theory of
Fields (Pergamon Press, 1975).

\bibitem{2}Hans Stephani, Relativity (Cambridge University Press, 2004).

\bibitem{3}Robert M. Wald, General Relativity (The University of Chicago Press, 1984).

\bibitem{4}A. Einstein, Kritisches zu einer von Hrn. de Sitter gegebenen L$\ddot{\mathrm{o}}$sung der Gravitationsgleichungen (Sitz. Preuss. Akad. Wiss., 1918, 1, 270-272, 1918).

\bibitem{5}R. Penrose, Singularities and time-asymmetry, in: General relativity, An Einstein centenary survey, ed.: S.W. Hawking, W. Israel (Cambridge University Press, 1979).

\bibitem{6}A.D. Linde, Particle Physics and Inflationary Cosmology
(CRC Press, 1990), (``Nauka'', Moscow, in Russian, 1990).

\bibitem{7}Vladimir S. Mashkevich, The Eternal Closed Universe: Deflation-Inflation (arXiv: 0910.0894v2 [physics.gen-ph], 2009).

\bibitem{8}Vladimir S. Mashkevich, On Dark Matter Problem:
Pseudomatter---Concept and Applications, arXiv: gr--qc/0902.3508
(2009).



\end{thebibliography}
\end{document}